\documentclass[aps,10pt,prc,tightenlines,twocolumn,twoside,showpacs,nofootinbib]{revtex4}
\usepackage{graphicx,amsmath,amssymb}
\usepackage{hyperref}

\newcommand{\dd}{\mathrm{d}}

\newcommand{\gsim}{\gtrsim}

\begin{document}

\title{Medium Modifications and Production of Charmonia at LHC}

\author{Xingbo Zhao$^1$ and Ralf Rapp$^2$} 
\affiliation{
$^1$ Department of Physics and Astronomy, Iowa State University, Ames, Iowa
50011, USA \\
$^2$Cyclotron Institute 
  and Department of Physics and Astronomy, Texas A\&M University, 
  College Station, TX 77843-3366, USA}

\date{\today}

\begin{abstract}
A previously constructed transport approach to calculate the evolution
of quarkonium yields and spectra in heavy-ion collisions is applied to 
Pb-Pb($\sqrt{s}$=2.76\,ATeV) collisions at the Large Hadron Collider 
(LHC). In this approach spectral properties of charmonia are constrained
by euclidean correlators from thermal lattice QCD and subsequently
implemented into a Boltzmann equation accounting for both suppression
and regeneration reactions. Based on a fair description of SPS and
RHIC data, we provide predictions for the centrality dependence of
$J/\psi$ yields at LHC. The main uncertainty is associated with the
input charm cross section, in particular its hitherto unknown reduction
due to shadowing in nuclear collisions. Incomplete charm-quark
thermalization and non-equilibrium in charmonium chemistry entail
a marked reduction of the regeneration yield compared to the statistical
equilibrium limit.  

\end{abstract}


\maketitle

\section{Introduction}
\label{sec_intro}
In recent years it has been realized that early ideas of associating
charmonium suppression in ultrarelativistic heavy-ion collisions 
(URHICs) with the deconfinement transition~\cite{Matsui:1986dk} are less 
direct than originally hoped for, see 
Refs.~\cite{Rapp:2008tf,Kluberg:2009wc,BraunMunzinger:2009ih} for
recent reviews. While there is little doubt that a sufficiently 
hot and dense medium created in URHICs induces a large suppression of 
primordially (or would-be) produced charmonia, coalescence processes of 
charm and anti-charm quarks in the later stages of the medium evolution 
are now believed to play an important role in a quantitative description 
of charmonium observables. The regeneration reactions are, in fact, 
dictated by the principle of detailed balance and are operative once
a charmonium bound state can be supported in the heat 
bath~\cite{Grandchamp:2003uw,Thews:2005vj,Yan:2006ve,Zhao:2007hh,Young:2008he,Linnyk:2008uf,Song:2010er}. 
The onset of regeneration for each charmomium state in the cooling process 
of the bulk matter is thus dictated by its  ``dissociation temperature", 
$T_{\rm diss}^\Psi$, which in general will be different for ground 
($\Psi$=$\eta_c$,$J/\psi$) and excited states ($\Psi$=$\psi'$,$\chi_c$, 
etc.). Furthermore, the inelastic reaction rate, $\Gamma_\Psi$,
of each charmonium controls its approach to the equilibrium limit. Both 
reaction rate (inelastic width) and equilibrium limit depend on the 
temperature and thus change as the system evolves in a heavy-ion 
reaction. The balance between primordially produced and regenerated 
charmonia can therefore be expected to be sensitive to the 
spectral properties of charmonia in the Quark-Gluon Plasma (QGP).  

In Ref.~\cite{Zhao:2010nk} we have implemented basic properties of  
charmonium spectral functions (mass, width and binding energy) into
a Boltzmann transport equation. The spectral properties have been
estimated from an in-medium potential model~\cite{Riek:2010fk} 
and checked against euclidean correlators from thermal lattice QCD.
Since the definition of an in-medium two-body heavy-quark potential 
remains an open problem to date, we have investigated two limiting
cases representing a ``weak-binding scenario" (WBS) and a 
``strong-binding scenario" (SBS) which are characterized by a small 
(1.2\,$T_c$) and large (2\,$T_c$) $J/\psi$ dissociation temperature,
respectively. The corresponding phenomenological analysis of $J/\psi$ 
production in URHICs at SPS ($\sqrt{s}$=17.3\,AGeV) and 
RHIC ($\sqrt{s}$=200\,AGeV), solving a rate equation in a 
thermally expanding fireball background, revealed that both SBS and WBS 
can be compatible with currently available data, despite the factor
of $\sim$10 difference in collision energy. The key difference is the 
{\em composition} of the final yield, especially in central $A$-$A$ 
collisions, where the regeneration (primordial) component is much 
larger in the WBS (SBS) than in the SBS (WBS). However, since the 
individual excitation functions ($\sqrt{s}$ dependence) of the 
primordial and regenerated components are quite opposite (decreasing 
vs. increasing, respectively)~\cite{Grandchamp:2001pf}, it is hoped
that another factor of $\sim$10 increase in collision energy is 
able to disentangle at least the two limiting scenarios.     
An initial investigation of this possibility is the main purpose
of this paper, by providing predictions for inclusive $J/\psi$ 
production in Pb-Pb collisions at $\sqrt{s}$=2760\,AGeV as recently
conducted at the Large Hadron Collider (LHC). Another important point 
is in how far chemical and thermal off-equilibrium effects cause a
deviation from the equilibrium limit as predicted, e.g., in the 
statistical hadronization model~\cite{Andronic:2006ky,Andronic:2010dt}.

The main uncertainty in our predictions will be associated with the input
for the open-charm cross section (as well as its shadowing in Pb-Pb).
Here, preliminary LHC data from $p$-$p$ collisions will be utilized
in connection with theoretical estimates for shadowing effects.
The charged particle multiplicity, $dN_{ch}/dy$, has already been 
measured and will be used to constrain the total entropy in the
system at each centrality. The total multiplicity is not only important 
for the initial temperature and its evolution but also determines the 
volume of the system at given temperature which affects the charm-quark 
density (for given total number of charm pairs, $N_{c\bar c}$) and thus 
the equilibrium limit of charmonia.  

In the following we first recapitulate our framework for
calculating the evolution of charmonia in URHICs and the connection
to their spectral properties (Sec.~\ref{sec_rate-eq}), 
discuss the open- and hidden-charm input quantities for
Pb-Pb at $\sqrt{s}$=2760\,AGeV (Sec.~\ref{sec_input}), 
evaluate results for the centrality dependence of inclusive $J/\psi$
yields (Sec.~\ref{sec_yields}) and conclude (Sec.~\ref{sec_concl}).

\section{Rate Equation and Spectral Properties of Charmonia}
\label{sec_rate-eq}
In this section we briefly recall the main features of constructing the 
equilibrium properties of charmonia ($\Psi$=$J/\psi$, $\chi_c$ and 
$\psi'$) in the Quark-Gluon Plasma (QGP) and hadron gas (HG), and their 
implementation into a thermal rate equation following 
Ref.~\cite{Zhao:2010nk} (which contains a more detailed account of
our approach). Upon integrating over the spatial and momentum dependence
of the Boltzmann equation one arrives at a simplified (averaged) 
rate equation for the number, $N_\Psi(\tau)$, of state $\Psi$ in a 
heavy-ion collision,  
\begin{equation}
\frac{\dd N_{\Psi}}{\dd \tau}=
-\Gamma_{\Psi}(T) \ [N_{\Psi}-N_{\Psi}^{\rm eq}(T) ] \ .
\label{rate-eq}
\end{equation}
The dissociation (and formation) rate, $\Gamma_\Psi$, is computed
in the so-called quasifree approximation for the processes
$p+\Psi \to p+ c+ \bar c$ induced by light partons, $p$, of the
heat bath ($u$, $d$, $s$ anti-/quarks and gluons)~\cite{Grandchamp:2001pf}.  
Besides the trivial $T$ dependence induced by the number density
of the light partons, the charmoniun binding energy,
\begin{equation}
\varepsilon_B(T)=2m^*_c(T)-m_{\Psi}(T) \ ,  
\label{epsB}
\end{equation}
plays an important in that states with larger 
$\varepsilon_B$ are characterized by a smaller $\Gamma_{\Psi}(T)$;
this ensures a realistic hierarchy between ground and excited states,
and is responsible for the main difference in $\Gamma_{\Psi}(T)$ 
between the SBS and WBS. The binding energies, $\varepsilon_B(T)$, 
in these two scenarios are taken from $T$-matrix 
calculations~\cite{Riek:2010fk} using lQCD results for either 
the internal ($U_{Q\bar Q}(r;T)$) or free energy ($F_{Q\bar Q}(r;T)$)   
as interaction potential, respectively. Within each scenario the 
in-medium charm-quark mass is fixed according to 
$2m^*_c(T) = U_{Q\bar Q}(\infty;T)$ or $F_{Q\bar Q}(\infty;T)$, which
also determines the charmonium mass, $m_{\Psi}(T)$, via Eq.~(\ref{epsB}).
The latter figures into the statistical equilibrium limit of each state, 
\begin{equation}
N_{\Psi}^{\rm stat}(T)=\gamma_c^2(N_{c\bar c};T) \ V_{\rm FB} \ 
d_{\Psi} \int\frac{d^3p}{(2\pi)^3} f^\Psi(p;T) \ , 
\label{Npsi-stat}
\end{equation}
where $d_{\Psi}$ denotes the spin degeneracy, $f^\Psi(p;T)$ the Bose 
distribution and $V_{\rm FB}$ the (time-dependent) fireball volume. The 
charm-quark fugacity, $\gamma_c(T)=\gamma_{\bar c}(T)$, is determined 
by assuming a fixed number, $N_{c\bar c}$, of charm-quark pairs in the
fireball for a given collision centrality, as following from primordial
production in binary $N$-$N$ collisions (the underlying $c\bar c$
cross section is one of the main input parameters discussed in 
Sec.~\ref{sec_input} below). For a given $V_{\rm FB}$ and temperature, 
$\gamma_c$ is calculated from relative chemical equilibrium among
the open- and hidden-charm states in the system, where the former
are either charm hadrons in the HG phase (using vacuum masses) or charm 
quarks with mass $m_c^*(T)$ in the QGP phase (charmonia have a negligible 
impact on $\gamma_c$ at the temperatures of interest here).  
To account for incomplete thermalization of the charm-quark spectra in 
the QGP, we adopt a relaxation-time approximation to correct the 
statistical equilibrium limit of charmonia~\cite{Grandchamp:2002wp}. 
The equilibrium limit figuring into the rate equation (\ref{rate-eq}) 
thus takes the final form
\begin{equation}
N_{\Psi}^{\rm eq}={\mathcal R}(\tau) \  N_{\Psi}^{\rm stat} \ , \
{\mathcal R}(\tau)=1-\exp (-\tau/\tau^{\rm eq}_c) \ . 
\end{equation}
The thermal relaxation time $\tau^{\rm eq}_c$ has been taken as a fit
parameter which controls the regeneration contribution in the rate 
equation. The resulting values, $\tau^{\rm eq}_c\simeq$\,2-4\,fm/c
for the WBS and SBS, respectively, are on the same order as microscopic 
calculations of charm-quark transport which lead to a fair description of
heavy-quark observables at RHIC~\cite{vanHees:2007me}. The impact of 
the ${\mathcal R}$-factor on the $J/\psi$ regeneration yield at SPS
and RHIC is appreciable, and quite comparable to a schematic coalescence 
model where the limiting cases of thermalized and initial power-law
spectra have been studied~\cite{Greco:2003vf}. Incomplete charm-quark
thermalization will also play a significant role in our predictions for 
LHC as discussed below.   

To check the equilibrium properties of charmonia against lQCD ``data", 
we have ``reconstructed" spectral functions by combining a relativistic 
Breit-Wigner ansatz for the bound-state part (using $M_\Psi(T)$ and 
$\Gamma_\Psi(T)$ as described above, as well as a polestrength factor 
$Z_\Psi(T)$ to mimic the wave-function overlap at the origin) with a 
nonperturbative continuum (with threshold $2m_c^*(T)$). Pertinent
euclidean correlator ratios turn out to be rather stable with $T$,
deviating from 1 by ca.~$\pm$10\% for both SBS and WBS, roughly 
consistent with lQCD. The vanishing of $Z_\Psi(T)$ is used to 
quantify the dissociation temperature, $T_{\rm diss}^\Psi$, above
which the gain term in the rate equation is switched off.  

\section{Input cross sections at $\sqrt{s}$=2760\,GeV}
\label{sec_input}
In this section we collect the inputs required for the initial
conditions of the rate equation in the charm and charmonium sector,
as well as for the thermal fireball evolution.  

For the total open-charm cross section in $p$-$p$ collisions at 
$\sqrt{s}=5.5$~TeV recent calculations using next-to-leading order (NLO) 
perturbative QCD (pQCD) obtain 2.5-3.5\,mb~\cite{Vogt:1900zz} using 
$m_c$=1.5\,GeV, compared to 5-7.5\,mb in earlier calculations with 
$m_c$=1.2\,GeV~\cite{Bedjidian:2004gd}. Preliminary LHC data
at 7~TeV~\cite{Dainese:2010ms} indicate a cross section close to the 
upper end of the pQCD predictions, which is around 10\,mb. Reducing 
this range by ca.~1/3-1/2 to extrapolate to $\sqrt{s}=2.76$~TeV, we estimate
$\sigma_{pp}^{c\bar c}(\sqrt{s}=2.76{\rm TeV})\simeq$\,5-8\,mb, in 
agreement with fixed-order-next-to-leading-log (FONLL) pQCD
predictions~\cite{Cacciari:2011priv,Cacciari:2005rk}. Converting 
this to a midrapidity density~\cite{Bedjidian:2004gd} (which amounts to
dividing by a factor of 
7-8~\cite{Cacciari:2011priv,Cacciari:2005rk,Bedjidian:2004gd}), we 
arrive at $\frac{d\sigma_{pp}^{c\bar c}}{dy}(\sqrt{s}=2.76\,{\rm TeV})
\simeq$\,0.7-1\,mb. Our calculations reported below are therefore 
conducted with a default value of 
$\frac{d\sigma_{pp}^{c\bar c}}{dy}(\sqrt{s}=2.76\,{\rm TeV})=0.75$\,mb
(which leads to a reasonable $(J/\psi)/(c\bar c)$ ratio when
evaluating available $J/\psi$ cross sections, see below).
The extrapolation to Pb-Pb collisions is done via standard binary-collision
($N_{\rm coll}$) scaling using an inelastic $p$-$p$ cross section of 65~mb 
accounting. At LHC energies the nuclear modification of parton distribution
functions in the nucleon (shadowing) is predicted to suppress the charm 
cross section significantly. We will investigate this effect by 
performing calculations where, according to Ref.~\cite{Bedjidian:2004gd},
the input charm cross section is reduced by up to 1/3 for 
central Pb-Pb, with a centrality dependence as estimated
from Refs.~\cite{Tuchin:2007pf}.

For the charmonium cross section $p$-$p$ measurements around midrapidity
are available from CDF ($\sqrt{s}$=1.96\,TeV)~\cite{Acosta:2004yw} and 
ALICE ($\sqrt{s}$=7\,TeV)~\cite{Dainese:2010ms} with 
$\frac{d\sigma_{pp}^{J/\psi}}{dy}=(3.4\pm0.3)\mu b$ and $(7.5\pm 2)\mu b$, 
respectively. In addition, charm measurements at lower energies 
suggest an approximately constant (energy-independent) fraction of 
$J/\psi$ to open-charm cross sections, 
$\frac{d\sigma_{pp}^{J/\psi}}{dy}/\frac{d\sigma_{pp}^{c\bar c}}{dy}\simeq
(0.5-1)$~\%, see, e.g., Ref.~\cite{Andronic:2006ky}. Thus we take 
$\frac{d\sigma_{pp}^{J/\psi}}{dy}(\sqrt{s}=2.76~{\rm TeV})=4~\mu b$,
which, with the above charm cross section, implies a $J/\psi$-to-$c\bar c$ 
fraction of 0.53\%. When including charmonium shadowing in Pb-Pb we
assume a suppression by 1/3 in central collisions, consistent with recent 
estimates at $\sqrt{s}$=5.5\,ATeV at midrapidity~\cite{Vogt:2010aa} using 
the color-evaporation model; this is somewhat smaller than the factor of 0.5
suppression found in Ref.~\cite{Abreu:2007kv} using EKS98~\cite{eks98} or 
nDSg~\cite{nDSg} shadowing, or a suppression by 2/3 or more predicted by 
the Color Glass Condensate approach~\cite{Kharzeev:2008nw,Tuchin:2010pv}.
Cold-Nuclear-Matter (CNM) effects other than shadowing are neglected.
Recent empirical evaluations of the nuclear absorption cross section in 
$p$-$A$ collisions indicate a marked reduction with increasing $\sqrt{s}$, 
suggesting it to be very small in the LHC energy regime. This is naively 
expected since the passage time of the nuclei at LHC has become very
small. We also neglect the Cronin effect. An initial-state $p_t$
broadening could be induced by shadowing effects which are expected 
to be reduced with increasing $p_t$, but we neglect this in the 
present work. It turns out that our results for the $J/\psi$ nuclear 
modification factor are not sensitive to CNM effects on the $J/\psi$ 
except for rather peripheral collision where the primordial component 
is sizable.

\begin{figure}[!t]
\centering
\includegraphics[angle=0,width=0.48\textwidth]{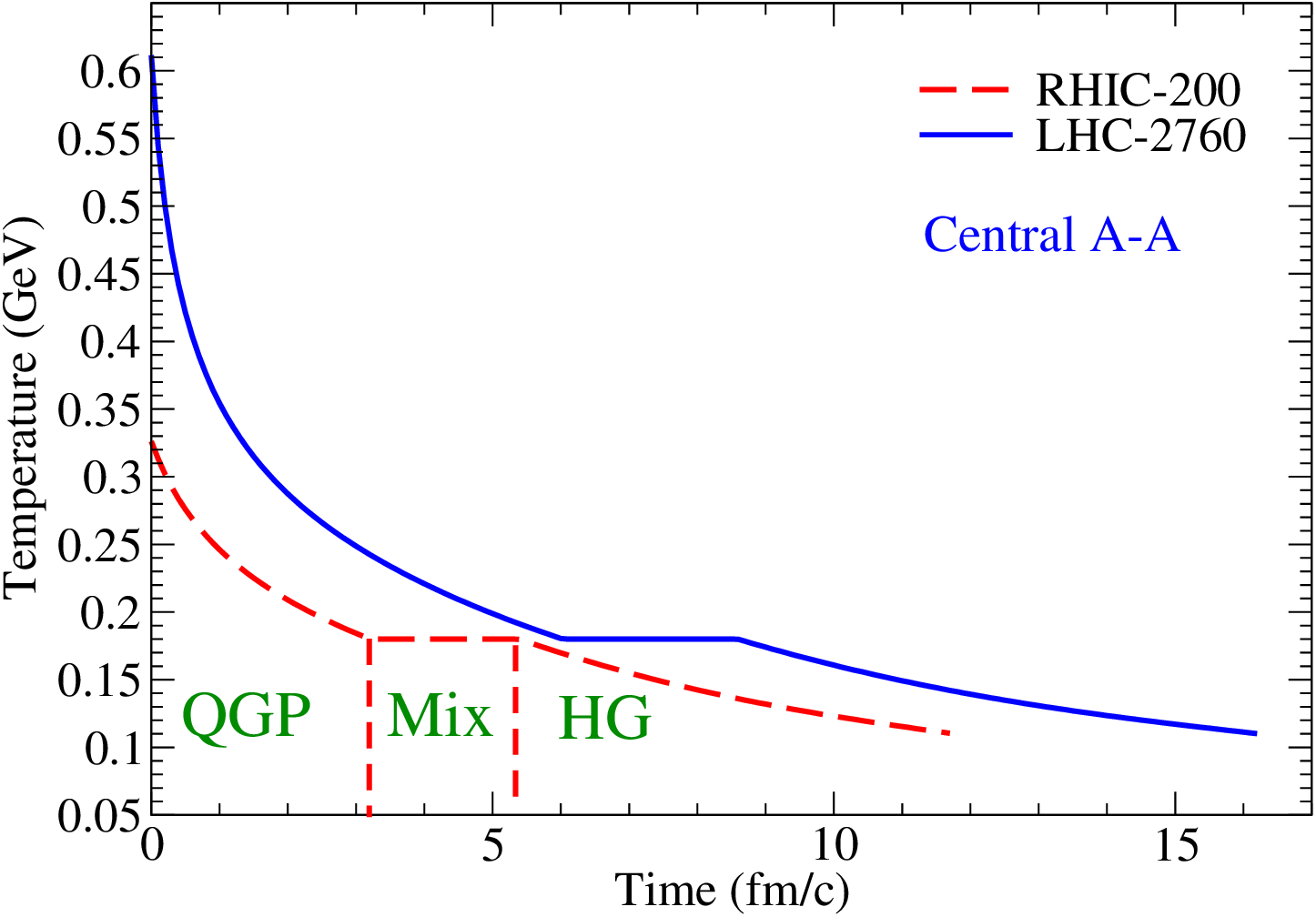}
\caption{(Color online) Time profile of temperature in central ($b$=0) 
 Au-Au ($\sqrt{s}$=0.2\,ATeV, $N_{\rm part}\simeq$~380; dashed line) and 
 Pb-Pb ($\sqrt{s}$=2.76\,ATeV, $N_{\rm part}\simeq$~400; solid line) 
 collisions using empirical charged-hadron multiplicities, 
 $dN_{\rm ch}/d\eta\simeq750$ and 1750, respectively, as input.} 
\label{fig_fb}
\end{figure}

For the thermal medium evolution, the main empirical input is the total
entropy, $S$, in the fireball which we assume to be conserved. We 
calculate the entropy density using a hadron-resonance gas equation of 
state (EoS) at chemical freezeout 
($s_{\rm chem}^{\rm HG}\simeq 6$\,fm$^{-3}$ with 
$T_{\rm chem}=180$~MeV). The 3-volume is then adjusted to obtain a 
charged-hadron multiplicity of $dN_{\rm ch}/d\eta\simeq1600$ for 0-5\% 
central Pb-Pb  at $\sqrt{s}$=2.76\,ATeV 
($N_{\rm part}\simeq$~380)~\cite{Aamodt:2010pb}, 
and the measured centrality dependence is employed for non-central 
collisions~\cite{Aamodt:2010cz}. Using a QGP quasiparticle 
EoS for $T>T_c=T_{\rm chem}=180$~MeV, we connect the two phases via a 
first-order transition, which allows to construct the temperature 
evolution, $T(\tau)$,  based on a cylindrical isotropic fireball volume 
expansion, $V_{FB}(\tau)$, using $s(T)=S/V_{FB}(\tau)$.  With a 
formation time of $\tau_0=0.2$\,fm/$c$ the initial temperature
amounts to $T_0=610$\,MeV, compared to 330\,MeV at RHIC. The lifetime 
of the QGP phase (where most of the $J/\psi$ chemistry is operative)
is also significantly larger, $\tau_{QGP}\simeq6$\,fm/$c$ vs. 3\,fm/$c$  
at RHIC, see Fig.~\ref{fig_fb}. 
  
\section{Centrality dependence of $J/\psi$ yields in 
Pb-Pb($\sqrt{s}$=2760\,GeV)}
\label{sec_yields}
With charmonium and open-charm input cross sections, as well as the 
fireball evolution at each centrality fixed, we solve the rate equation 
(\ref{rate-eq}) for the 3 charmonia $\Psi=J/\psi$, $\psi'$ and $\chi_c$ 
and obtain the final inclusive $J/\psi$ yield using standard feeddown 
fractions, i.e., 32\% and 8\% from $\psi$' and $\chi_c$ in $pp$ collisions,
respectively. In addition we account for $B$-meson feeddown (Bfd) 
according to Tevatron data~\cite{Acosta:2004yw}, amounting to ca.~10\% 
of the inclusive $J/\psi$ yield in $pp$ (without any modifications 
of its $p_t$ dependence, i.e., $b$-quark energy loss), see, e.g., 
Ref.~\cite{Zhao:2008vu}. 
We display the $J/\psi$ yields in terms of the 
nuclear modification, either as a function of collision centrality
(characterized by the number of nucleon participants, $N_{\rm part}$),
\begin{equation}
R_{AA}(N_{\rm part}) = \frac{N_{J/\psi}(N_{\rm part})}
{N_{\rm coll}(N_{\rm part}) \  N_{J/\psi}^{pp}} \ , 
\label{raa-npart}
\end{equation}
or at fixed centrality as a function of fireball evolution time,
\begin{equation}
R_{AA}(\tau) = \frac{N_{J/\psi}(\tau)}
{N_{\rm coll}(N_{\rm part}) \  N_{J/\psi}^{pp}} \ ,
\label{raa-t}
\end{equation}
or transverse momentum, 
\begin{equation}
R_{AA}(p_t) = \frac{dN_{J/\psi}/dp_t}
{N_{\rm coll}(N_{\rm part}) \  dN_{J/\psi}^{pp}/dp_t} \ .
\label{raa-pt}
\end{equation}

\begin{figure}[!t]
\centering
\includegraphics[angle=0,width=0.465\textwidth]{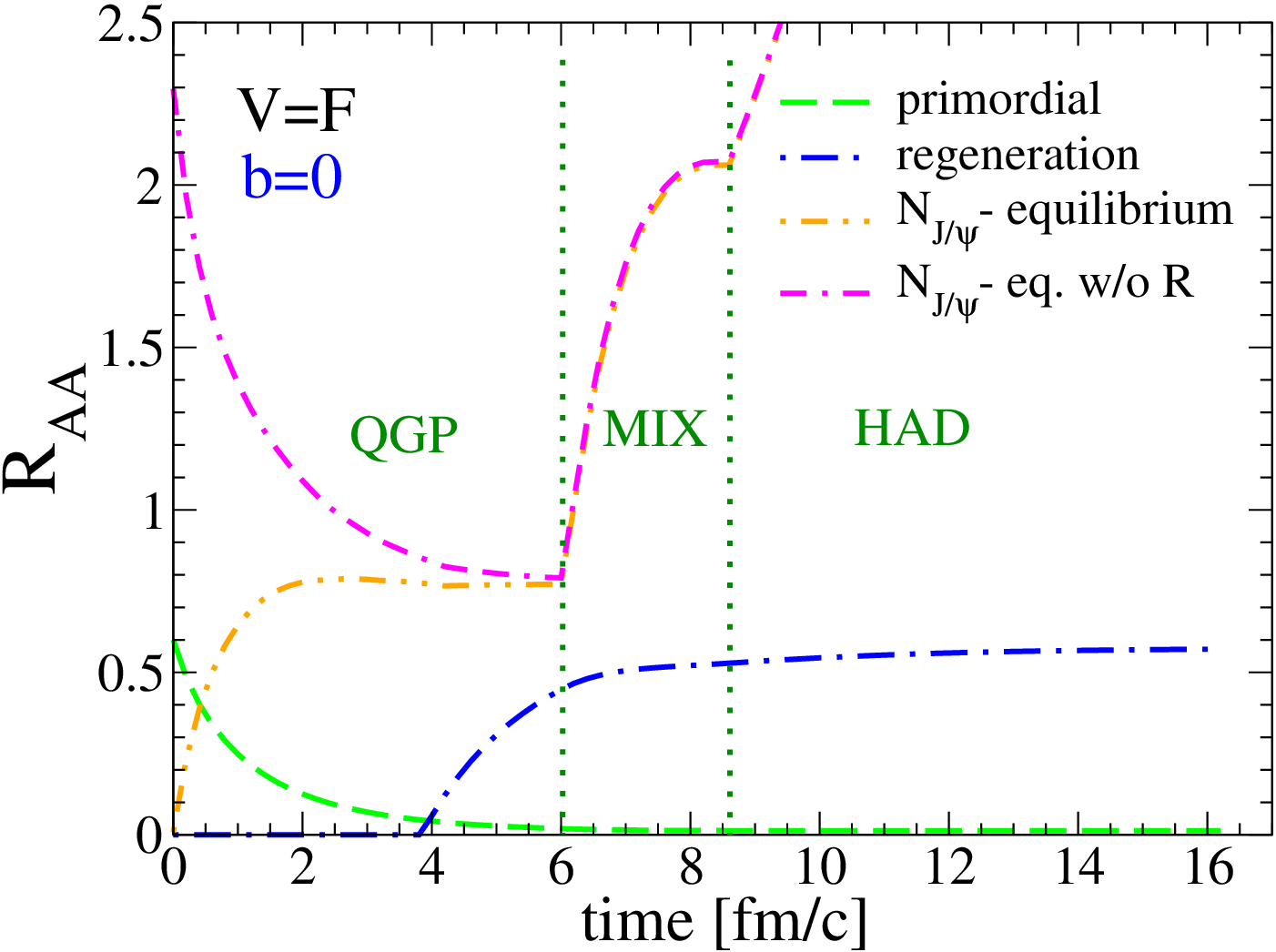}
\includegraphics[angle=0,width=0.465\textwidth]{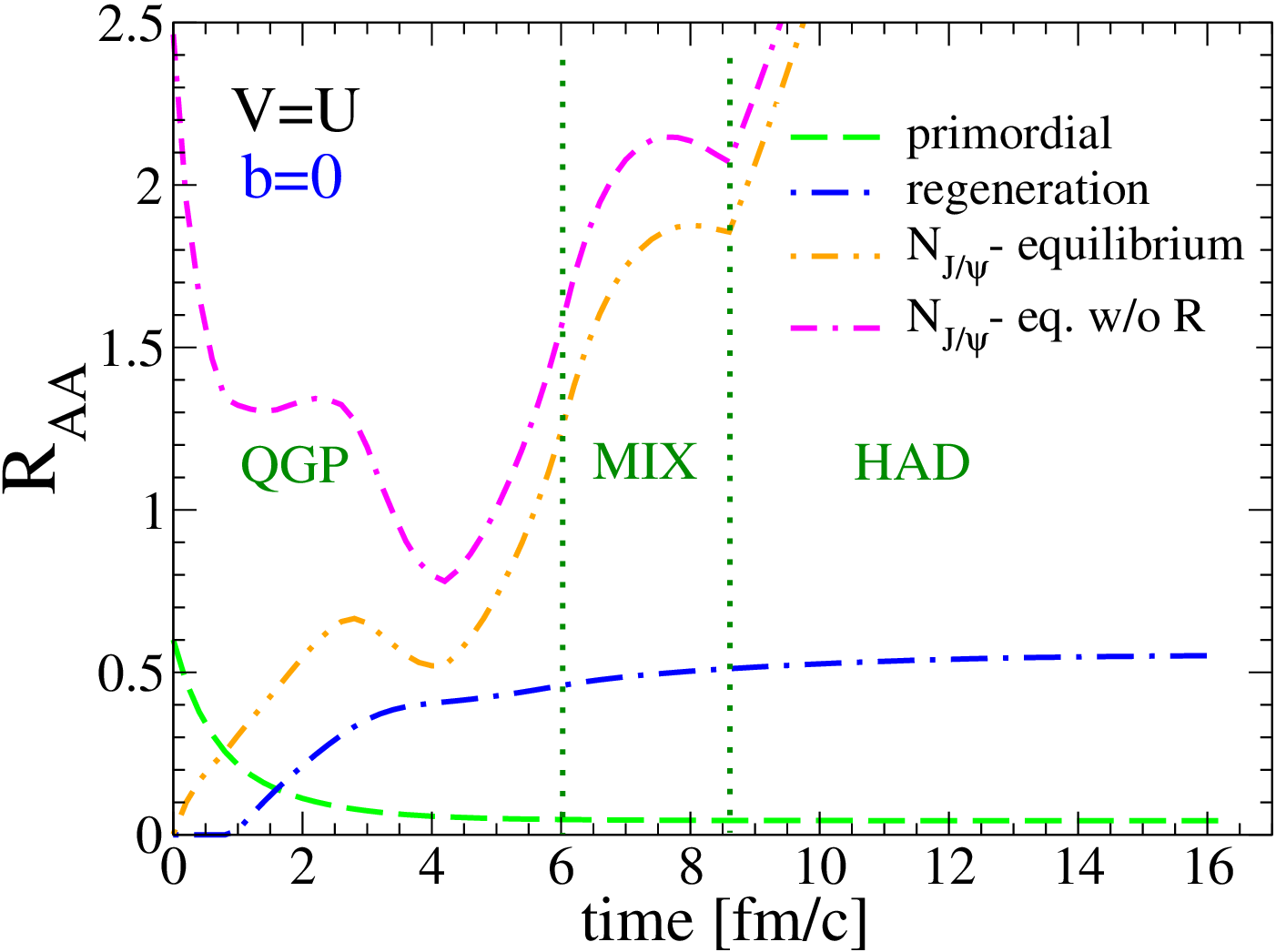}
\caption{(Color online) Time dependence of the $J/\psi$ nuclear modification
factor in central ($b$=0) Pb-Pb collisions at $\sqrt{s}$=2.76\,ATeV 
(participant number $N_{\rm part}$\,$\simeq$\,400)
at $\sqrt{s}$=2.76\,ATeV within the WBS (top) and SBS (bottom). No shadowing
effects are included; only the exclusive $J/\psi$ contribution is included
in the numerator of $R_{AA}(t)$, Eq.~(\ref{raa-t}), while its denominator 
includes the 40\% feeddown fraction, to facilitate the comparsion with 
subsequent plots (consequently, $R_{AA}$($\tau$=0)=0.6). Bottom feeddown and
formation-time effects are not included.} 
\label{fig_time}
\end{figure}
Let us first examine the time dependence of the exclusive $J/\psi$
yield as a function of fireball evolution time, displayed in
Fig.~\ref{fig_time} for central Pb-Pb in the weak- and strong-binding
scenario (upper and lower panel, respectively), using
$\frac{\sigma_{pp}^{c\bar c}}{dy}=0.75$\,mb without shadowing.
In both scenarios the large initial temperature leads to a strong and 
rapid suppression of the primordially produced $J/\psi$'s. The 
somewhat surprising feature is that the final regeneration yield is 
very similar in both scenarios, despite the different binding energies 
and dissociation rates at given temperature. The time dependence of
the regenerated component furthermore shows that the $c\bar c$
coalescence occurs at rather different times or, equivalently, 
temperature, namely in the regime where the gain term in the 
rate equation first becomes operative, i.e., for $T_{\rm diss}<1.2(2)T_c$, 
corresponding to $\tau \gsim 4 (1)$\,fm/$c$ in the WBS (SBS). It is precisely
in this regime where the dissociation (and thus formation) rate is large, 
facilitating a rather rapid approach of the $J/\psi$ abundance toward 
its equilibrium limit. However, the ``cooking" only lasts for a duration 
of ca.~$\Delta \tau\simeq 2-3$\,fm/$c$ in both scenarios, after which the 
reaction rate becomes small so that the yield stabilizes.   

\begin{figure}[!t]
\centering
\includegraphics[angle=0,width=0.48\textwidth]{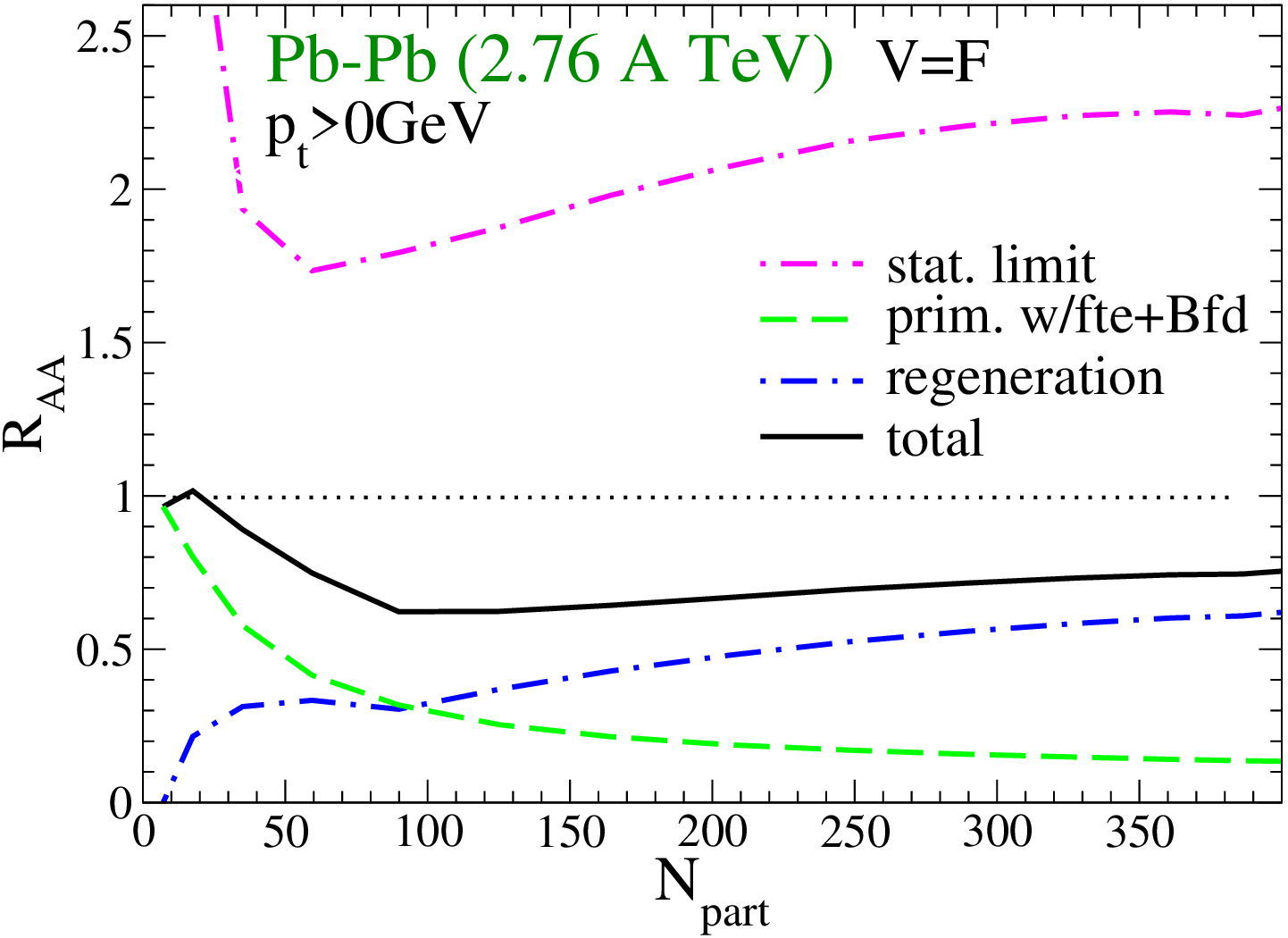}
\includegraphics[angle=0,width=0.48\textwidth]{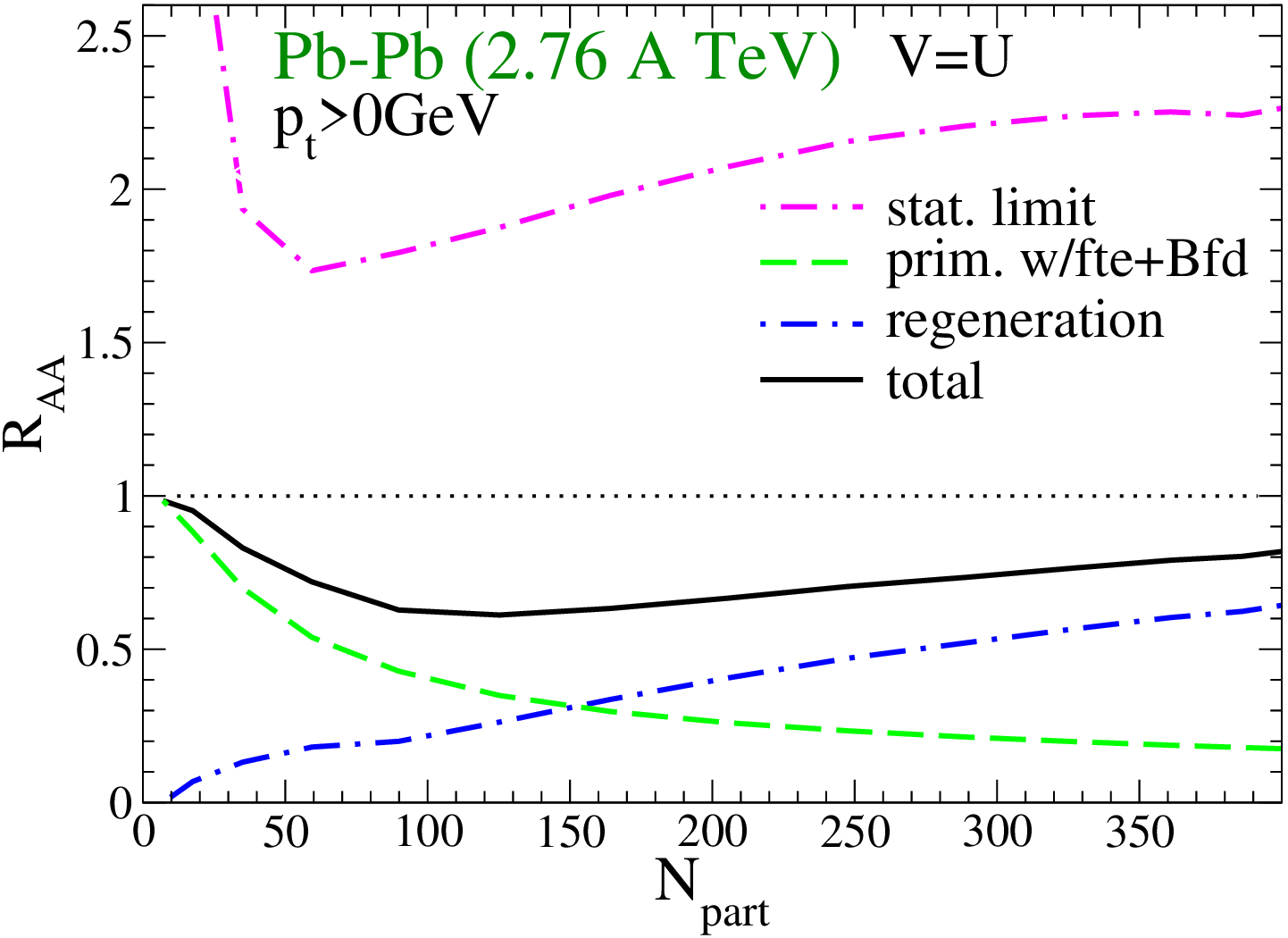}
\caption{(Color online) Centrality dependence of the nuclear modification 
factor, Eq.(\ref{raa-npart}),  for the inclusive $J/\psi$ yield (including 
feeddown) in Pb-Pb($\sqrt{s}$=2.76\,ATeV) collisions within weak- and 
strong-binding scenarios (upper and lower panel, respectively).}
\label{fig_raa}
\end{figure}
\begin{figure}[!t]
\centering
\includegraphics[angle=0,width=0.48\textwidth]{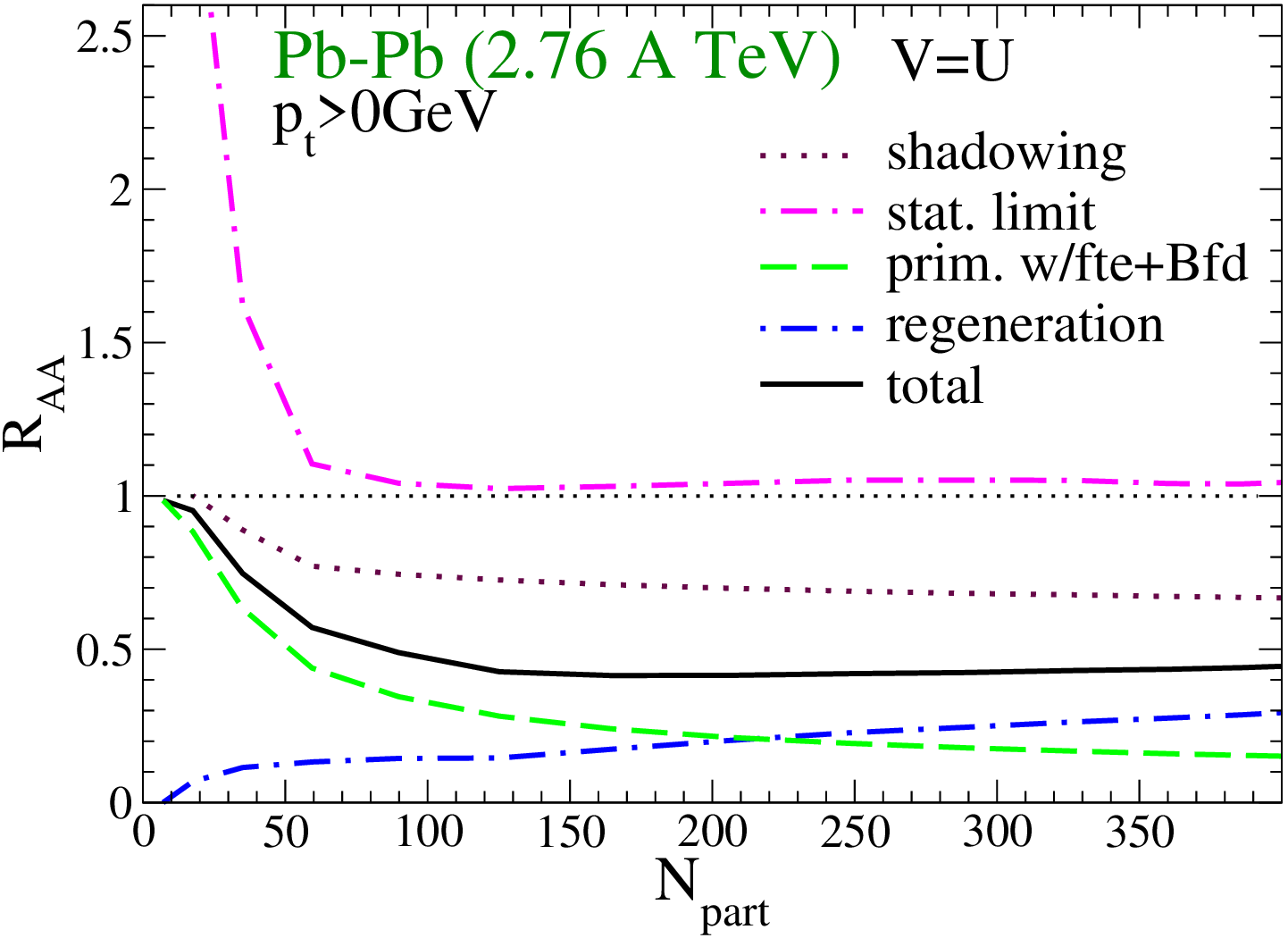}
\caption{(Color online) Nuclear modification factor for the inclusive
$J/\psi$ yield (including feeddown) in Pb-Pb($\sqrt{s}$=2.76\,ATeV)
collisions within the strong-binding scenario including a shadowing
correction of the $N_{\rm coll}$-scaled open-charm and $J/\psi$ cross
sections by a factor of 2/3 for $N_{\rm part}$\,$\simeq$\,400.}
\label{fig_raa-cc-sh}
\end{figure}
The centrality dependence of the final $J/\psi$ yields (including 
feeddown) in Pb-Pb collisions is displayed in Fig.~\ref{fig_raa}, again
for both WBS (upper panel) and SBS (lower panel) and for the same 
conditions as in the time dependence shown in Fig.~\ref{fig_time} 
(i.e, $N_{\rm coll}$-scaled charm cross section of
$\frac{\sigma_{pp}^{c\bar c}}{dy}=0.75$\,mb without shadowing),
except that bottom feeddown (Bfd) and formation-time effects (fte) 
are now included as detailed in Ref.~\cite{Zhao:2008vu}. One 
finds that the similarity of the two scenarios not only holds for 
central but also for semicentral and even peripheral collisions.
For $N_{\rm part}\simeq200$ the composition of the total $J/\psi$ 
yield in terms of primordial and regenerated components is quite
comparable to central Au-Au collisions at RHIC, as is the 
charged-particle multiplicity. However, due to a smaller initial 
overlap volume at LHC (at identical $dN_{\rm ch}/dy$) the initial 
temperature is significantly larger than at RHIC. 
In addition, charm production is larger and thus regeneration effects are
more pronounced. 

A striking feature of our predictions is a rather large 
deviation of the final $J/\psi$ yield from the values predicted by the 
statistical model of hadron production (upper dash-dotted lines in 
Fig.~\ref{fig_raa}). By definition, these abundances refer to an 
equilibrated hadron gas at hadrochemical freezeout (which is naturally 
identified with the hadronization transition), with a fixed
charm-quark number enforced by the fugacity, $\gamma_c$, in 
Eq.~(\ref{Npsi-stat}). The equilibrium limit, $N_\Psi^{\rm eq}$, in our 
rate equation (\ref{rate-eq}) basically coincides with the statistical 
equilibrium limit, $N_\Psi^{\rm stat}$, since the thermal 
off-equilibrium ${\cal R}$-factor has become close to one at the end of 
the mixed phase (recall Fig.~\ref{fig_time}). While the observed light- 
and strange-hadron abundances in central $AA$ collisions are in excellent 
agreement with the statistical model~\cite{BraunMunzinger:2003zd}, our 
calculations suggest that the inelastic charmonium reaction rates are 
too small in the vicinity of $T_c$, especially on the hadronic side, to 
establish relative chemical equilibrium in the charm/onium sector.
Consequently, the $J/\psi$ yields following from the kinetic approach 
are well below the statistical-model limit (by about a factor of $\sim$3),
resulting in a suppression below one in the nuclear modification factor.

As is well known, in the grand canonical limit the equilibrium $\Psi$ 
number depends quadratically on the number of open-charm pairs in the
system, $N_\Psi^{\rm eq}\propto \gamma_c^2 \propto N_{c\bar c}^2$.
Therefore, regeneration (or statistical production) are quite sensitive
to the open-charm cross section. In Fig.~\ref{fig_raa-cc-sh} we display 
our results for $R_{AA}^{J/\psi}(N_{\rm part})$ in the SBS including a 
($p_t$-independent) shadowing correction to  both 
$N_{c\bar c}(b)=N_{\rm coll}(b) \ N_{c\bar c}^{pp}$ and $N_{\Psi}(b)$,
as described in Sec.~\ref{sec_input}. As 
expected, the reduction of the $N_{c\bar c}$ by 1/3 in central Pb-Pb 
suppresses the regeneration $J/\psi$ yield by a factor of 
ca.~$(2/3)^2\simeq 1/2$, so that the total $R_{AA}$ decreases from 
$\sim$0.8 to 0.45 (no shadowing is applied to the bottom feeddown 
contribution).  
This underlines again the importance of an accurate open-charm input
cross section for in-medium charmonium physics at LHC.

\begin{figure}[!t]
\centering
\includegraphics[angle=0,width=0.48\textwidth]{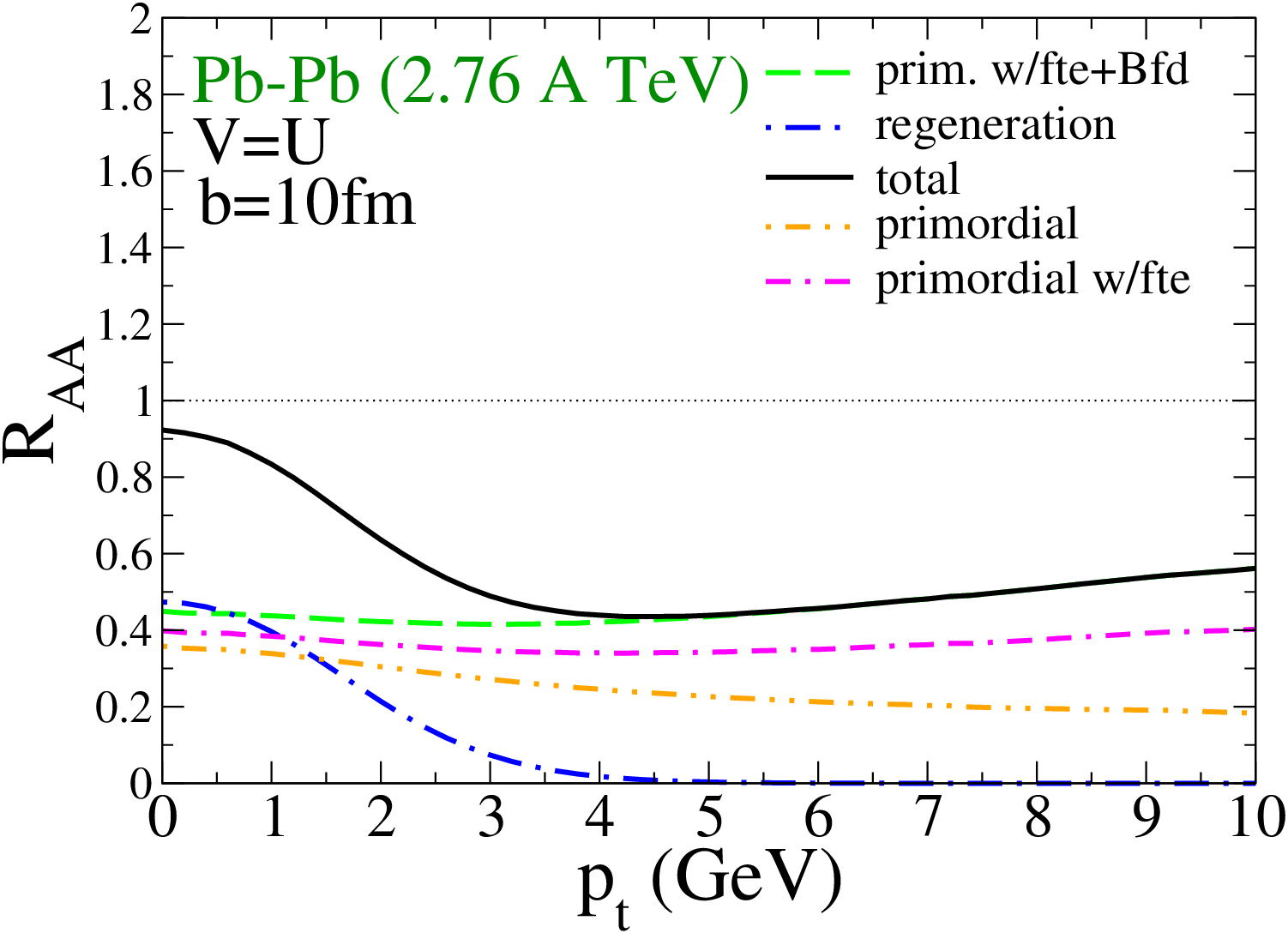}
\includegraphics[angle=0,width=0.48\textwidth]{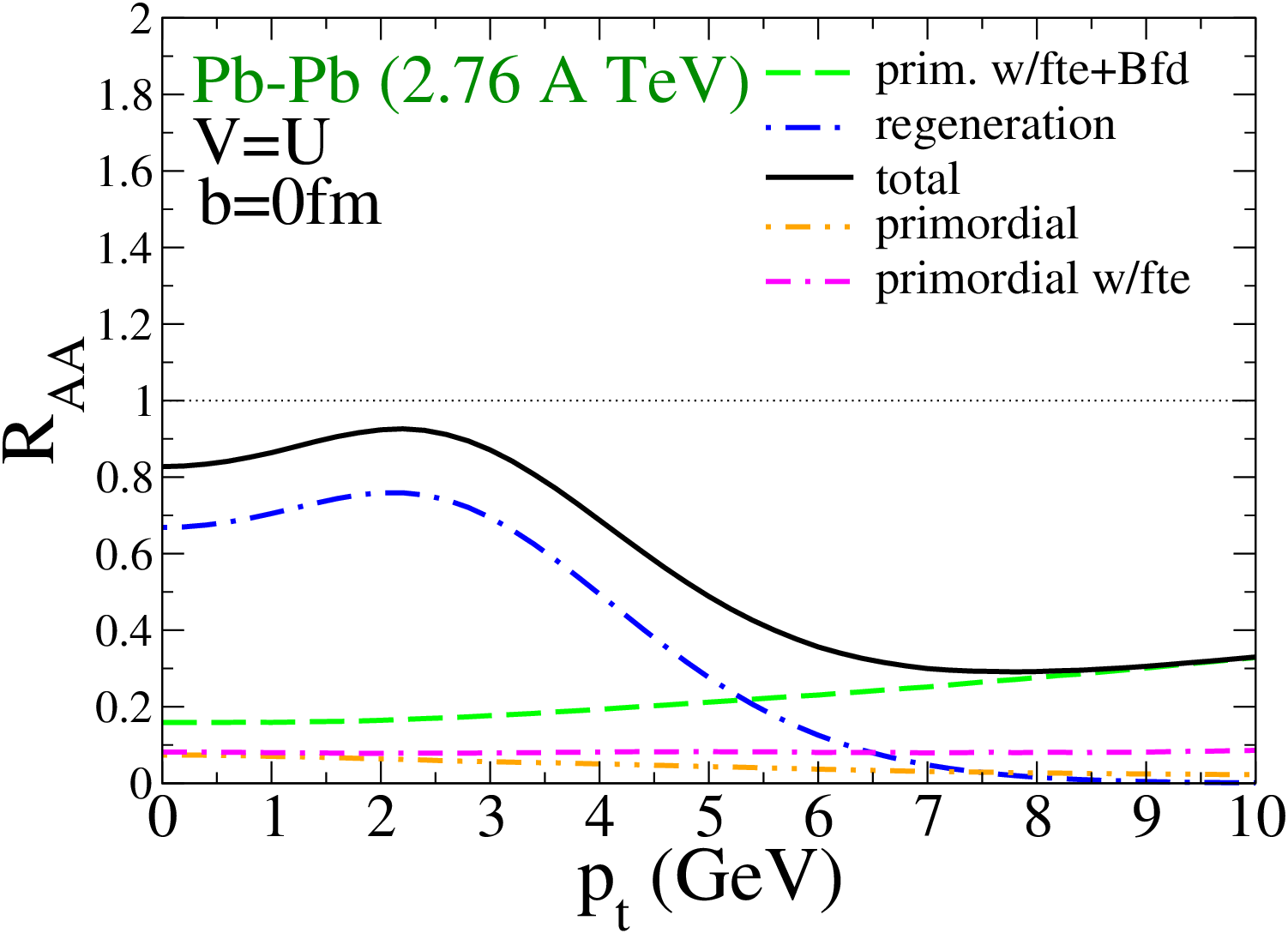}
\caption{(Color online) Transverse-momentum dependence of the nuclear
modification factor, Eq.(\ref{raa-pt}), for inclusive $J/\psi$'s 
(including feeddown) in peripheral (upper panel) and central (lower 
panel) Pb-Pb($\sqrt{s}$=2.76\,ATeV) collisions within the strong-binding 
scenario (no shadowing corrections accounted for).}
\label{fig_raa-pt}
\end{figure}
Next, we investigate the transverse-momentum spectra of inclusive
$J/\psi$ production, displayed in Fig.~\ref{fig_raa-pt} for peripheral 
and central Pb-Pb collisions (upper and lower panel, respectively). For 
simplicity we restrict ourselves to predictions without shadowing 
corrections. At both centralities the regeneration yield generates a 
marked enhancement of the spectra at low $p_t$. The maximum structure 
is more pronounced in central collisions due to larger regeneration and 
larger suppression of the primordial production; the crossing between 
the two components occurs at $p_t\simeq$\,5-6\,GeV, compared to 
$p_t\simeq$\,0\,GeV for peripheral collisions. The latter is reminiscent
of central Au-Au collisions in the SBS at RHIC~\cite{Zhao:2010nk}, 
although there no significant maximum develops at low $p_t$ (the charm 
cross section is too small and the suppression of the primordial 
component is stronger in central Au-Au($\sqrt{s}$=0.2\,ATeV) than in 
peripheral Pb-Pb($\sqrt{s}$=2.76\,ATeV).
We note, however, that shadowing corrections are expected to be most 
pronounced at low $p_t$ (i.e., small values of 
$x=2p_t/\sqrt{s}$ at midrapidity) and fade away at high $p_t$. 
Since shadowing suppresses both regeneration and primordial production,
our calculations in Fig.~\ref{fig_raa-pt} may be viewed as an upper
limit for the maximum at low $p_t$.

\begin{figure}[!t]
\centering
\includegraphics[angle=0,width=0.48\textwidth]{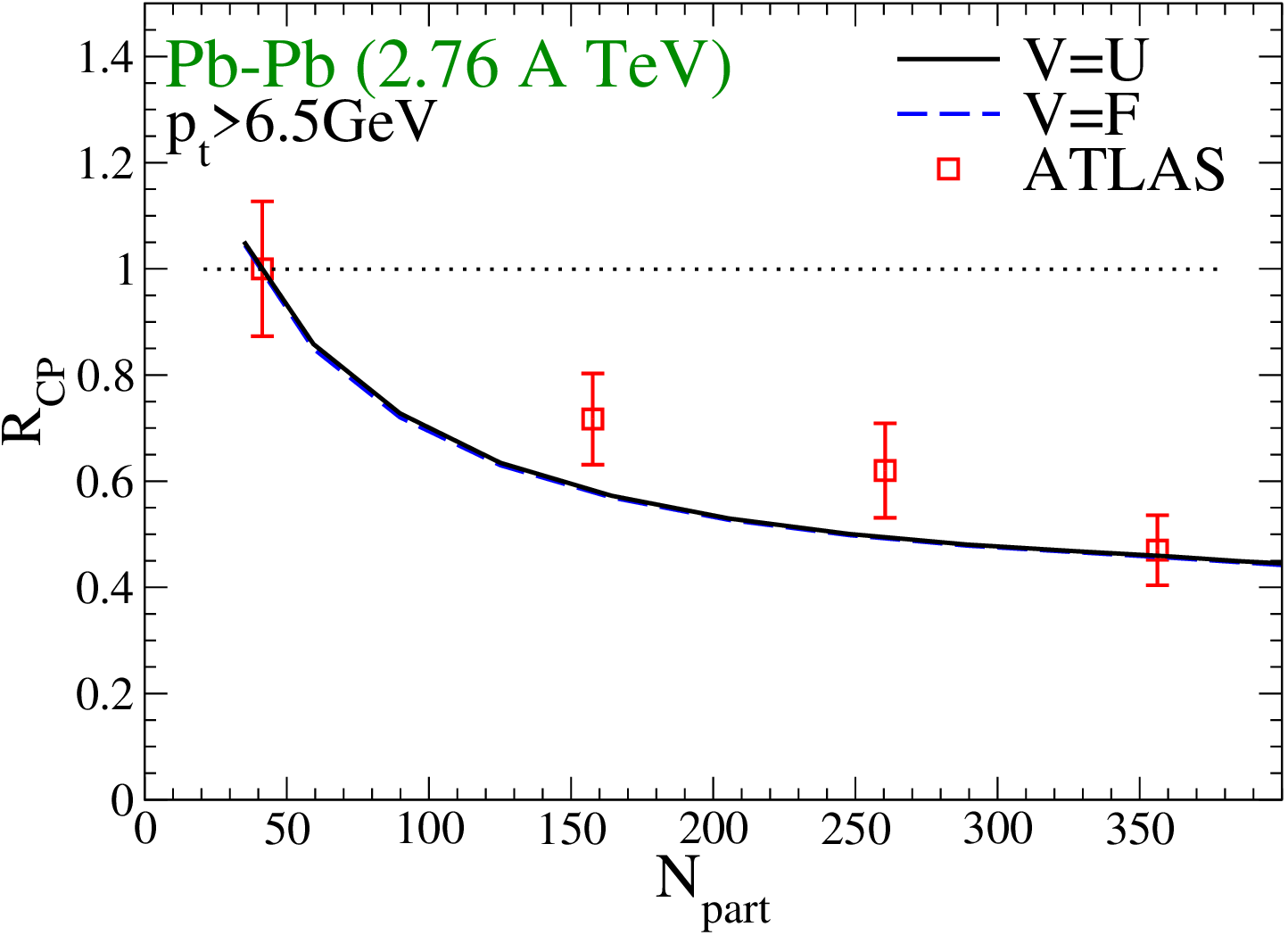}
\caption{(Color online) Central-to-peripheral ratio, $R_{CP}$, for inclusive
$J/\psi$ yields (including feeddown) in Pb-Pb($\sqrt{s}$=2.76\,ATeV)
collisions within the strong- and weak-binding scenario (solid and dashed 
line, respectively; no shadowing included). A $p_t>6.5$\,GeV cut has been 
applied to all $J/\psi$'s in the calculations to compare
to ATLAS data~\cite{Aad:2010px}.}
\label{fig_rcp}
\end{figure}

Finally, we  perform a calculation for the so-called central-to-peripheral
ratio of $J/\psi$ mesons, which is similar to $R_{AA}$ but normalized
to a peripheral centrality bin,  
\begin{equation}
R_{CP}(N_{\rm part}) = 
   \frac{N_\psi(N_{\rm part})/N_{\rm coll}(N_{\rm part})}
        {N_\psi^{\rm peri}/ N_{\rm coll}^{\rm peri}} \ . 
\end{equation}
A first measurement of this quantity in Pb-Pb at LHC has recently been
reported by the ATLAS collaboration~\cite{Aad:2010px} where the 
peripheral bin corresponds to a 40-80\% centrality selection. In 
addition, only muons with $p_t^\mu>3$~GeV were accepted, implying that 
ca.~80\% of the reconstructed $J/\psi$ carry a transverse momentum of 
$p_t>6.5$\,GeV. We mimic these conditions in our calculation by 
representing the 40-80\% centrality bin with an average 
$N_{\rm part}\simeq40$ and using a momentum cut of $p_t>6.5$\,GeV
for all $J/\psi$'s (neglecting shadowing).  
The latter implies that the regeneration contribution is essentially
gone, recall Fig.~\ref{fig_raa-pt} so that our calculated $R_{CP}$ only 
involves primordially (suppressed) $J/\psi$'s, including bottom feeddown 
and formation-time effects in the QGP suppression 
reactions~\cite{Zhao:2008vu}. The trend of our postdiction (without any 
adjustment of parameters) is fairly compatible with the 
ATLAS measurement, cf.~Fig.~\ref{fig_rcp}. We emphasize again that 
the high-$p_t$ cut evades conclusions about regeneration mechanism
other than that they are not operative in this regime.  

\section{Conclusion}
\label{sec_concl}
Employing a kinetic rate-equation approach which incorporates in-medium
effects extracted from charmonium spectral functions we have computed 
inclusive $J/\psi$ production in Pb-Pb collisions as recently conducted 
at the LHC. Our approach reproduces existing charmonium data at SPS and 
RHIC energies where the two main model parameters had been adjusted 
(strong coupling constant in the dissociation rate and thermal 
off-equilibrium correction in the equilibrium limit). Our pertinent
predictions for LHC indicate a large suppression of primordial 
charmonia (by a factor of $\sim$10 in central Pb-Pb) and thus a 
predominance of the regeneration yield for participant numbers 
$N_{\rm part}>100-150$. However, the final yield stays well below the 
predictions of the statistical model, by about a factor of $\sim$3, 
implying an $R_{AA}^{J/\psi}$ below one even for an input charm cross 
section of 0.75\,mb in $pp$ collisions; shadowing corrections on this 
quantity further suppress the final $J/\psi$ yield. Somewhat 
surprisingly, both strong- and weak-binding scenarios give very similar 
results, which we attribute to the rather limited time window 
($\Delta \tau\simeq 2-3$\,fm/$c$) during which the gain term in the rate 
equation is active (below $T_{\rm diss}$) and the inelastic reaction 
rate is large. In the strong-binding scenario $J/\psi$'s form at 
significantly higher temperatures and thus at earlier times in the 
fireball evolution than for weak binding.
This may lead to discernible differences in the finally observed
elliptic flow, which, however, will have to be measured with rather
good precision. We also performed calculations of $R_{CP}^{J/\psi}$ in 
comparison to very recent ATLAS data, finding a suppression roughly 
compatible with experiment. However, due to the large-$p_t$ cut in 
these data ($p_t>$\,6.5\,GeV), our result is not sensitive to regeneration 
contributions which only figure at lower $p_t\lesssim 6(3)$\,GeV for
central (peripheral) collisions. 
Upcoming LHC data are eagerly awaited to test our predictions and
give further insights into charmonium production mechanisms in
hot/dense QCD matter.

\acknowledgments 
We thank A.~Andronic, M.~Cacciari, E.~Musto,  J.~Schuhkraft and K.~Tuchin 
for valuable discussions. This work is supported by the US National Science 
Foundation under grant no. PHY-0969394 (RR, XZ),  by the A.-v.-Humboldt 
foundation (RR) and by the US Department of Energy under grant no. 
DE-FG02-87ER40371 (XZ).

\end{document}